\documentclass[showpacs,showkeys,12pt,aps]{revtex4}
\usepackage{graphicx}
\usepackage{hyperref}
\usepackage{latexsym}
\usepackage{amssymb}
\usepackage{color}

\begin{document}
\title{Higher dimensional generalization of Buchdahl-Vaidya-Tikekar model for super compact star}

 \author{Avas Khugaev}\email{avaskhugaev@mail.ru }
 \affiliation{Institute of Nuclear Physics,
 Tashkent, 100214, Uzbekistan}

 \author{Naresh Dadhich}\email{nkd@iucaa.in}
\affiliation{Centre for Theoretical Physics, Jamia Millia, Islamia, New Delhi, 110025, India, Inter-University Center for Astronomy and Astrophysics, Post Bag 4 Pune 411 007, India}

 \author{Alfred Molina}\email{alfred.molina.ub.edu}
  \affiliation{Departament de F\'\i sica Fonamental, Institut de Ci\'encies del Cosmos,
 Universitat de Barcelona, Spain}


\begin{abstract}
We obtain higher dimensional solutions for super compact star for the Buchdahl-Vaidya-Tikekar metric ansatz. In particular, Vaidya and Tikekar characterized the $3$-geometry by a parameter, $K$ which is related to the sign of density gradient. It turns out that the key pressure isotropy equation continues to have the same Gauss form, and hence $4$-dimensional solutions can be taken over to higher dimensions with $K$ satisfying the relation, $K_n = (K_4-n+4)/(n-3)$ where subscript refers to dimension of spacetime. Further $K\geq0$ is required else density would have undesirable feature of increasing with radius, and the equality indicates a constant density star described by the Schwarzschild interior solution. This means for a given $K_4$, maximum dimension could only be $n=K_4+4$, else $K_n$ will turn negative. 
\end{abstract}
\maketitle

\section{Introduction}

There exists a vast literature on relativistic models for a star interior. The simplest and the oldest one is the constant density sphere described by Schwarzschild interior solution, and its universality for higher dimensions as well as for Lovelock gravity has also been established \cite{Naresh1}. Another interesting fluid solution is isothermal sphere with a linear equation of state $p=\gamma\rho \sim 1/r^{n-2}$ where $n$ is dimension of spacetime. It turns out that it also has universal solution for pure Lovelock (for which action and the equation of motion has only one term corresponding to $N$, the degree of Lovelock action) gravity \cite{dhm}. The former has undesirable feature of sound velocity being infinite, yet it accords to typical star interior parameters and overall behaviour while for the latter density is singular at the center though mass goes to zero, and it describes a relaxed equilibrium state of distribution without compact boundary. It could very well approximate a star or galaxy cluster. A physically reasonable distribution with a finite boundary lies between these two extremes. 

There is a fairly general metric ansatz due to Buchdahl \cite{Buchdahl} that covers almost all interesting solutions. Vaidya and Tikekar \cite{Vaidya} particularized Buchdahl ansatz by giving it a geometric meaning. They envisioned that $3$-hypersurface is embedded in $4$ dimensional spheroid rather than sphere, and it is characterized by a dimensionless parameter $K$, a measure of deviation from sphericity. Interestingly it turns out that sign of this parameter is also associated with density gradient. That is for gradient to be non-positive, $K$ must be non-negative and $K=0$ gives constant density. This indicates for a given mass or radius, star is densest when density is constant. 

The point to be appreciated is that besides sphere and spheroid one can explore various topologies and see whether they accord to some meaningful physical distributions? For instance, if we consider a torus instead of spheroid, then we will write  
\begin{equation}
(\sqrt{x^2+y^2+z^2}-L)^2+ w^2 =R^2
\label{eqn_41}
\end{equation}
where $L$ is distance from torus center to tube center and $R$ is tube radius. This will lead to the metric,
$$
ds^2=e^{\nu}dt^2-\frac{R^2}{R^2-(r-L)^2} dr^2-r^2 d\Omega^{2}_{2} 
$$
describing a fluid distribution with diverging density at center. It is therefore not physically acceptable. 

Vaidya and Tikekar obtained solution for the case of $K=2$ \cite{Vaidya} and the most general solution for integer values of $K$ was obtained by Mukherjee, Paul and Dadhich \cite{Naresh}. It may be noted that this metric ansatz is particularly suited for describing compact objects like neutron stars \cite{knutsen}. In this paper we wish to extend Buchdahl-Vaidya-Tikekar (BVT) model for compact star to higher dimensions. It turns out that the key pressure isotropy equation has the same Gauss form in higher dimensions as well. Hence $4$-dimensional solutions could be carried over to higher dimensions with appropriately redefining the parameter $K$ which satisfies the relation $K_n = (K_4-n+4)/(n-3)$ where subscript indicates spacetime dimension. It may however be noted that  though the solution is the same yet distribution is not entirely identical because density and pressure depend upon the parameter $K$ which is different for different $n$. For a given $K_4$, there would be a spectrum of $K_n$ values for different $n$ which has a cut-off at $n=K_4+4$ when $K_n=0$ representing uniform density.  

There is also fairly large literature on star interior solution in higher dimensions, we give some representative references \cite{Koikawa1, Koikawa2, Krori1, Krori2, Maharaj, Patel, Sudan}. We have here studied how $4$-dimensional solution could be taken over to  appropriate higher dimensions in BVT metric ansatz. The paper is organized as follows: In the next section we write BVT metric and set up the equations for perfect fluid distribution. In Sec 3, we discuss the generalized higher dimensional solutions followed by discussion of their physical features.We end up with a discussion. 

\section{Metric and Einstein equations}
Let us begin with the general static spherically symmetric metric in the $n$--dimensional spacetime,
\begin{eqnarray}
ds^2=e^{\nu}dt^2-e^{\lambda}dr^{2}- r^{2}d\Omega^{2}_{n-2}\,.
\label{eqn_2}
\end{eqnarray}

Substituting the metric in the Einstein equation for a perfect fluid distribution, we obtain density and pressure (see for instance, Ref. \cite{Naresh1} with  $\tilde\alpha=0$) as
\begin{equation}
\rho=\frac{n-2}{2r^2}e^{-\lambda}\Bigl(r\lambda^{\prime} - (n-3)(1-e^{\lambda})\Bigr)
\label{eqn_25}
\end{equation}
\begin{equation}
p=\frac{n-2}{2r^2}e^{-\lambda}\Bigl(r\nu^{\prime} + (n-3)(1-e^{\lambda})\Bigr)
\label{eqn_26}
\end{equation}
and the pressure isotropy equation is given by 
\begin{eqnarray}
\nu^{''}+\frac{1}{2}\nu^{'2}-
\biggl(\frac{1}{2}\lambda^{'}+\frac{1}{r}\biggr)\nu^{'} -
2(n-3)\biggl[\frac{1}{r}\biggl(\frac{1}{2}\lambda^{'}+\frac{1}{r}
\biggr)-\frac{e^{\lambda}}{r^{2}}\biggr]=0\,.
\label{eqn_1}
\end{eqnarray}
By writing $\psi=e^{\frac{1}{2}\nu}$, we rewrite the above equation 
\begin{eqnarray}
\psi^{\prime\prime}-
\biggl(\frac{1}{2}\lambda^{'}+\frac{1}{r}\biggr)\psi^{\prime} -
(n-3)\biggl[\frac{1}{r}\biggl(\frac{1}{2}\lambda^{'}+\frac{1}{r}
\biggr)-\frac{e^{\lambda}}{r^{2}}\biggr]\psi = 0 \,.
\label{eqn_3}
\end{eqnarray}

The Buchdahl ansatz \cite{Buchdahl} prescribes 
\begin{equation}
e^{\lambda} = \frac{1+cr^2}{1+(c-a)r^2}
\end{equation}
with $a>0, c>0$,
which was particularized by Vaidya and Tikekar \cite{Vaidya} to write
\begin{eqnarray}
e^{\lambda}=\frac{1+K\alpha r^{2}}{1-\alpha r^{2}}\,.
\label{eqn_4}
\end{eqnarray}
where $\alpha = R^{-2}$. Now density reads as follows: 
\begin{eqnarray}
\rho (r)= (K+1)(n-2)\alpha\frac{n-1 +(n-3)K\alpha r^2}{2(1+K\alpha r^2)^{2}}.
\label{eqn_27}
\end{eqnarray}
Clearly $K+1>0$ for density to be positive. For physical reasonableness, density gradient must be negative or zero,   
\begin{equation}
\rho(r)'=-K(K+1)(n-2)\alpha^2 r \frac{n+1+(n-3)K\alpha r^2}{(1+K\alpha r^2)^3} \leq 0.
\end{equation}

This requires $K>0$. The physical requirement of density decreasing with radius constrains the nature of $3$-geometry or deviation from sphericity parameter $K$ cannot be negative.  

Putting in the metric ansatz, the pressure isotropy equation takes the form
\begin{eqnarray}
(1-\alpha r^{2})(1+K\alpha r^{2})\psi^{\prime\prime}
-r^{-1}(1+K)\Bigl[1+
\frac{K}{1+K}(1-\alpha r^{2})^2\Bigr]\psi^{\prime}-\nonumber \\
-(n-3)(1+K)K\alpha^{2}r^{2}\psi =0
\label{eqn_7}
\end{eqnarray}
We further, as in Ref. \cite{Vaidya}, transform $r$ to 
\begin{eqnarray}
u^{2}=\frac{K}{K+1}(1-\alpha r^{2}),\quad K>0\nonumber \\
\label{eqn_8}
\end{eqnarray}
then Eqn ({\ref{eqn_3}) becomes
\begin{eqnarray}
(1-u^{2})\frac{d^{2}\psi}{du^{2}}+u\frac{d\psi}{du}+(1+K)(n-3)\psi=0 .
\label{eqn_11}
\end{eqnarray}
The only difference with the expression given in \cite{Vaidya} is the factor
$(n-3)$ in the last term of Eqn (\ref{eqn_11}). This second order equation has two singular regular points in $u=\pm 1$ and we can write the solution around $u=0$ as in:\cite{Vaidya}
\begin{eqnarray}
\psi=\sum_{m}A_{m}u^{m}
\label{eqn_12}
\end{eqnarray}
where now coefficients are determined in terms of $A_0$ and $A_1$, the two arbitrary constants of integration, as
\begin{eqnarray}
A_{m+2}=\frac{m^{2}-2m-(K+1)(n-3)}{(m+1)(m+2)}A_{m}.
\label{eqn_13}
\end{eqnarray}
Here even and odd coefficients are given respectively in terms of $A_0$ and $A_1$. In the two tables below we 
give for $K=2$ explicitly even and odd coefficients, and they are: 

\begin{table}[h]
\caption{Comparison for even $A_{m}$  coefficients at $K=2$, for $m=0,2,4,6$} 
{\begin{tabular}{lll}
\toprule m & $A_{m} (n>4)$ & $A_{m} (n=4)$ \\
\colrule
0 & $A_{0}$ & $A_{0}$ \\
2 & $-\frac{3}{2}(n-3)A_{0}$ & $-\frac{3}{2}A_{0}$\\
4 & $\frac{3}{8}(n-3)^{2}A_{0}$& $\frac{3}{8}A_{0}$\\
6 & $\frac{(17-3n)(n-3)^{2}}{80}A_{0}$ & $\frac{1}{16}A_{0}$ \\
\botrule
\end{tabular}\label{ta1} }
\end{table}
\begin{table}[h]
\caption{Comparison for odd $A_{m}$  coefficients at $K=2$, for $m=1,3,5,7$} 
{\begin{tabular}{lll}
\toprule m & $A_{m} (n>4)$ & $A_{m} (n=4)$ \\
\colrule
1 & $A_{1}$ & $A_{1}$ \\
3 & $-\frac{3n-8}{6}A_{1}$ & $-\frac{2}{3}A_{1}$\\
5 & $\frac{(3n-8)(n-4)}{40}A_{1}$& $0$\\
7 & $-\frac{(3n-8)(n-4)(n-8)}{560}A_{1}$ & $0$ \\
\botrule
\end{tabular}\label{ta2} }
\end{table}

From Eqn (\ref{eqn_13}) it is clear that corresponding to a $K_4$ value for $n=4$, the solution is the same for some $K_n$ in $n$ dimension. The general relation is 
\begin{eqnarray}
n-3 = \frac{K_4+1}{K_n+1}
\label{eqn_16}
\end{eqnarray}
where subscript refers to dimension of spacetime. The solution is the same with $K$ replaced. This means there exists a higher dimensional fluid compact star with the same solution corresponding to every $4$ dimensional solution with proper $K$ value as given by the above relation. Note that coefficients of the corresponding series will be the same in terms of constants $A_{0}$ and $A_{1}$. 

Let us now consider some particular cases. For $K_4=2$, $K_6=0$ which means six dimensional fluid sphere is of constant density. For $n>6$, $K_n$ will turn negative and thereby implying density gradient being positive and hence untenable. For a given $K_4$, the upper bound for dimension $n = K_4+4$, else $K_n<0$ which is not physically  admissible. Thus for $K_4=2$, $n=6$ is only possible while for $K_4=7$ there could be $n= {5, 7, 11}$ respectively corresponding to $K_n = 3, 1, 0$, and so on. Uniform density marks the limiting case of maximum density and once that is reached, one cannot go any further in $n$. 

Here we have taken $K\in \mathbb{Z}$ but that is not necessary, it can be any positive value. Uniform density sphere has the universal solution in the Schwarzschild interior metric irrespective of dimension of spacetime as well as Einstein or Lovelock gravity theory \cite{Naresh1}.

To gain some more insight let us recall the Vaidya-Tikekar construction \cite{Vaidya}. We consider a spheroidal hypersurface in $4$-Euclidean space defined by

\begin{equation} 
 \frac{x^2+y^2+z^2}{R^2} + \frac{w^2}{b^2} = 1
\end{equation}

which will generate the metric 
\begin{equation}
 d\sigma^2 = \frac{1+K\alpha r^2}{1-\alpha r^2} dr^2 + r^2 d\Omega_2^2 
\end{equation}
where 
\begin{equation}
 K=\frac{b^2}{R^2} - 1.
\end{equation}

Now the relation (\ref{eqn_16}) can be written in the equivalent form, as:
\begin{eqnarray}
K_4+1=(K_n+1)(n-3)\to \frac{b_4}{R_4}=\sqrt{n-3}\frac{b_n}{R_n} .
\label{eqn_18}
\end{eqnarray}
This brings out how spheroidal parameters are related in the two cases.

Note that to obtain a polynomical solution instead of a series in Eqn (13),  the coefficient $A_{m+2}$
must vanish. For $n=4$ this condition implies $K=(m-1)^2-2$, where $m\geq 3$. Generalization of this simple rule in the case of Eqn (\ref{eqn_11}) for a generic dimension $n$ leads to the requirement $m^2-2m-(K+1)(n-3)=0$,
which should have solution for positive integer number $K$ and $n\geq 4$. This shows there is a richer structure in higher dimensions \footnote{Other reformulation is that: for $n=4$ 
the finite series will be for $K=L^2+2$, here $L=0,\pm 1,\pm 2,...$, from where follows, that $A_m=0$ for any $m>1+\sqrt{K+2}$; for $n>4$
$A_m=0$ for any $m>1+\sqrt{1+(n-3)(K+1)}$ where $L$ is an according integer such as, that $K=\frac{L^2-n+2}{n-3}$ is an integer positive number.}

\section{The Solution}

As in \cite{Naresh}, a general solution for $n=4$ is given in terms of the Gegenbauer functions while in \cite{Buchdahl} it is expressed as the hypergeometric functions. By redefining variable, we can also write our solution in a more general form using hypergeometric functions, which would also include Gegenbauer functions as a particular case. Note that the equation (\ref{eqn_11}), obtained for a spacetime of arbitrary dimension $n>4$, can be transformed to a Gauss type differential equation \cite{Gauss}, by a simple redefinition of the variable,  $u$: $z=\frac{1-u}{2}$. Then we can transform Eqn. (\ref{eqn_11}) to the standard form of the Gauss differential equation, written as:\cite{Slater}
\begin{eqnarray}
z(1-z)\frac{d^{2}\psi}{dz^2}+\Bigl(-\frac{1}{2}+z\Bigr)\frac{d\psi}{dz}+(1+K)(n-3)\psi=0 \, .
\label{eqn_19}
\end{eqnarray}
and the solution can be expressed as
\begin{eqnarray}
\psi (z) =A\,\,
{_{2}F_{1}}(a,b;-\frac{1}{2},z)+B\,z^{\frac{3}{2}}{_{2}F_{1}}(a+\frac{3}{2},b+\frac{3}{2};\frac{5}{2},z)
\label{eqn_22}
\end{eqnarray}
where $A$ and $B$ are an arbitrary integration constants, and $_{2}F_{1}(a,b;c,z)$ is the Gauss hypergeometric function in the usual notation \cite{Slater}. 

The Gauss hypergeometric function has the following parameters :
\begin{eqnarray}
a=-1\pm j ,\quad b=-2-a=-1\mp j,\quad c=-\frac{1}{2} 
\label{eqn_20}
\end{eqnarray}
where $j=\sqrt{1+(1+K)(n-3)}$. Recall the parameter $K$ is restricted to be non-negative, $K \geq 0$. Thus we have the solution in closed form in terms of the Gauss hypergeometric functions \footnote{here also $|z|=\frac{1}{2}|1-u|<1$}.

It is obvious, that Eqn. (\ref{eqn_19}) has two singular points \footnote{we don't consider here the singularity at $z\to\infty$}, and they are $z=0, 1$. Near $z=0$, the solution takes the form 
\begin{eqnarray}
\psi (z) =A\,\,
{_{2}F_{1}}(a,b;-\frac{1}{2},z)+B\,z^{\frac{3}{2}}{_{2}F_{1}}(a+\frac{3}{2},b+\frac{3}{2};
\frac{5}{2},z)
\label{eqn_23}
\end{eqnarray}
while for near $z=1$, it reads as 
\begin{eqnarray}
\psi (z) =A\,\,
{_{2}F_{1}}(a,b;-\frac{1}{2},1-z)+
B\,(1-z)^{\frac{3}{2}}{_{2}F_{1}}(-\frac{1}{2}-b,-\frac{1}{2}-a;\frac{5}{2},1-z) \, .
\label{eqn_24}
\end{eqnarray}

We have given above in Eqns (\ref{eqn_27}) and (\ref{eqn_26}) the expressions for density and pressure. Let us consider
\begin{equation}
p(r)=-\rho(r)+\frac{n-2}{2r}\Bigl\{(1-e^{-\lambda})'+e^{-\lambda}\nu'\Bigr\}
\label{eqn_28}
\end{equation}
and we can write 
\begin{equation}
\nu'=\alpha r\sqrt{\frac{K}{(K+1)(1-\alpha r^2)}}\frac{1}{\psi (z)}\frac{d\psi}{dz} .
\label{eqn_29}
\end{equation}
To determine the unknown integration constants $A$ and $B$ we should match $g_{00}$ and $g_{11}$ functions in (\ref{eqn_2}) with the exterior Schwarzschild solution, 
\begin{equation}
ds^2=f(r)dt^2-f^{-1}(r)dr^{2}- r^{2}d\Omega^{2}_{n-2},
\label{eqn_2900}
\end{equation}
where
\begin{equation}
f(r)=1-\frac{C_n}{r^{n-3}}\quad C_n=\pi^{-\frac{n-1}{2}}\Gamma\Bigl(\frac{n-1}{2}\Bigr)\frac{k^{2}_{n}\cdot M}{n-2}.
\label{eqn_2910}
\end{equation}
Here $k^{2}_{n}$ is higher dimensional gravitation constant and $M$ is mass of star \cite{Kanti},\cite{Tangherlini}. The star boundary is defined by vanishing pressure, $p=0$ on the surface and we determine the free parameters as 
\begin{eqnarray}
M\to M_n=\frac{\pi^{\frac{n-1}{2}}(n-2)}{k^{2}_{n}\Gamma\Bigl(\frac{n-1}{2}\Bigr)}\frac{(1+K)\alpha r^{n-1}_{0}}{1+K\alpha r^{2}_{0}}\nonumber,
\label{eqn_2920}
\end{eqnarray}
\begin{eqnarray}
A=\frac{c\cdot \frac{dF_2}{dz}-d\cdot F_2}{F_1\frac{dF_2}{dz}-F_2\frac{dF_1}{dz}}|_{z_0},\qquad
B=-\frac{c\cdot \frac{dF_1}{dz}-d\cdot F_1}{F_1\frac{dF_2}{dz}-F_2\frac{dF_1}{dz}}|_{z_0} \nonumber,
\label{eqn_2930}
\end{eqnarray}
where
\begin{eqnarray}
c=\sqrt{1-\frac{C_n}{r^{n-3}_{0}}}=\sqrt{\frac{1-\alpha r^{2}_{0}}{1+K\alpha r^{2}_{0}}} \nonumber,
\label{eqn_2940}
\end{eqnarray}
\begin{eqnarray}
d=2\sqrt{\frac{(K+1)(1+K\alpha r^{2}_{0})}{K}}\Bigl(\frac{\rho (r_{0})}{\alpha (n-2)}
-\frac{1+K}{(1+K\alpha r^{2}_{0})^2}\Bigr) \nonumber,
\label{eqn_2950}
\end{eqnarray}
\begin{eqnarray}
F_{1}(z) ={_{2}F_{1}}(a,b;-\frac{1}{2},z), \quad F_{2}(z)=z^{\frac{3}{2}}{_{2}F_{1}}(a+\frac{3}{2},b+\frac{3}{2};\frac{5}{2},z)\nonumber .
\label{eqn_2960}
\end{eqnarray}
The variable $z$ is expressed in terms of $r$ \footnote{for $z_0$ we have $z_{0}=\frac{1}{2}\Bigl( 1 - \sqrt{\frac{K(1-\alpha r^{2}_{0})}{K+1}} \Bigr)$}:
\begin{eqnarray}
z=\frac{1}{2}\Bigl( 1 - \sqrt{\frac{K(1-\alpha r^2)}{K+1}} \Bigr)\nonumber,
\label{eqn_2970}
\end{eqnarray}
where $r_0$ is the star radius. For  $0\leq r \leq r_{0}$, density and pressure are given by 
\begin{eqnarray}
\rho (r)=\frac{\alpha (1+K)(n-2)(n-1+(n-3)K\alpha r^2)}{2(1+K\alpha r^2)^2}\nonumber,
\label{eqn_2980}
\end{eqnarray}
\begin{equation}
p(r)=-\rho (r)+\frac{\alpha (1+K)(n-2)}{(1+K\alpha r^2)^2}+\frac{\alpha (n-2)}{2(1+K\alpha r^2)}
\sqrt{\frac{K(1-\alpha r^2)}{K+1}}\frac{A\frac{dF_1}{dz}+B\frac{dF_2}{dz}}{A\cdot F_1+B\cdot F_2}.
\label{eqn_2990}
\end{equation}

To compute the derivatives $\frac{dF_1}{dz}$ and $\frac{dF_2}{dz}$ we can use some properties of the Gauss hypergeometric functions and obtain for them the following expressions:
\begin{eqnarray}
\frac{dF_1}{dz}=-2ab\cdot{_{2}F_{1}}(a+1,b+1;\frac{1}{2},z)
\label{eqn_30}
\end{eqnarray}
\begin{eqnarray}
\frac{dF_2}{dz}=\frac{3}{2}z^{\frac{1}{2}}\cdot {_{2}F_{1}}(a+\frac{3}{2},b+\frac{3}{2};\frac{5}{2},z)+\nonumber\\
+\frac{2}{5}(a+\frac{3}{2})(b+\frac{3}{2})z^{\frac{3}{2}}\cdot {_{2}F_{1}}(a+\frac{5}{2},b+\frac{5}{2};\frac{7}{2},z) .
\label{eqn_31}
\end{eqnarray}

In Figs. 1 and 2 we plot normalized density, $\rho/\rho_0(r=0)$ and pressure for the two cases $K_4=7, 14$. Since pressure value as well as its variation is very small for the case of constant density for $n=11, 18$, its blow up is shown in Fig 3. Note that for $K_4=7$, we have $K_n = 7, 3, 1,0$ corresponding respectively to $n=4,5,7,11$, while for $K_4=14$ it is $K_n=14, 4, 2, 0$ and $n=4, 6, 8, 18$. Clearly density variation slows down as dimension increases until constant density is reached (Fig 1). In Fig 2 we see that central pressure decreases with increasing dimension and so does its variation. For constant density distribution pressure has very small value and so is its variation (Fig 3). It should be noted that for the case $K=0$, the variable $u$ (\ref{eqn_12}) turns vacuous and hence it should be considered separately from the isotropy equation (\ref{eqn_7}) itself.
\begin{figure}
\includegraphics[width=7cm]{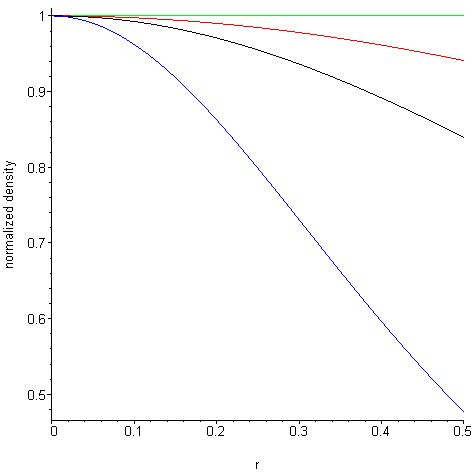}\hspace*{1cm}\includegraphics[width=7cm]{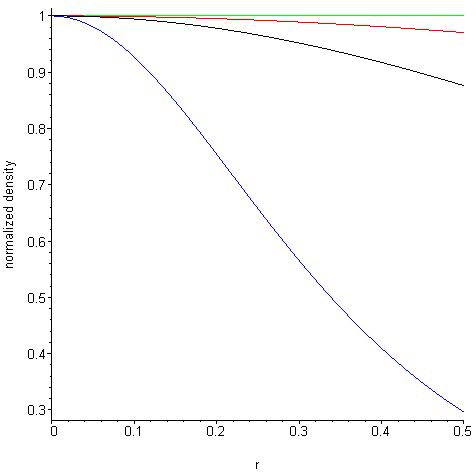}
\caption{\label{fig:density7} Normalized density plots show in ascending order the cases $K_4=7$ and $n=4, 5, 7, 11$ on left and while on right  $K_4=14$ and $n=4, 6, 8, 18$ respectively.}
\end{figure}

\begin{figure}
\includegraphics[width=7cm]{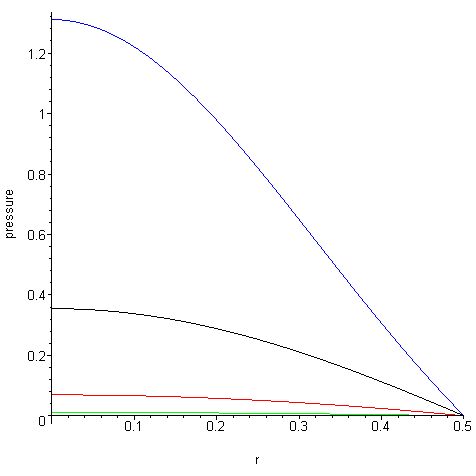}\hspace*{1cm}
\includegraphics[width=7cm]{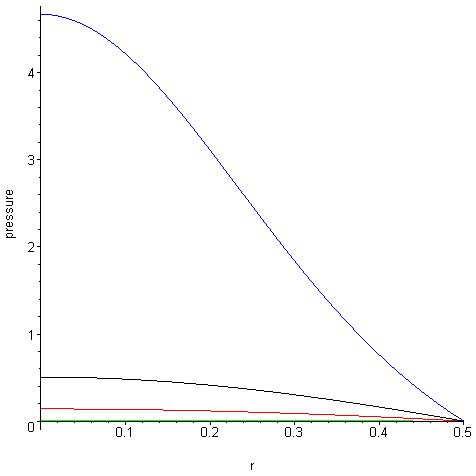}
\caption{\label{fig:pressure7} Pressure plots show in descending order the cases $K_4=7$ and $n=4, 5, 7, 11$ on left and while on right  $K_4=14$ and $n=4, 6, 8, 18$ respectively.}
\end{figure}

\begin{figure}
\includegraphics[width=7cm]{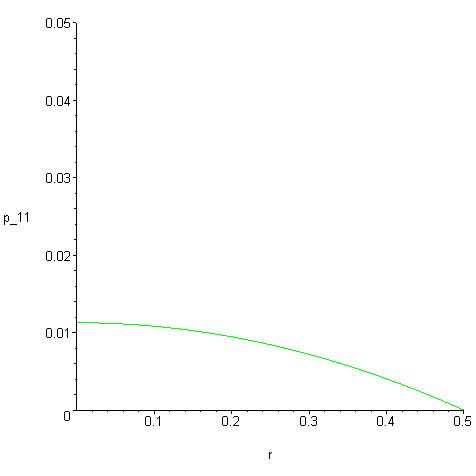}\hspace*{1cm}\includegraphics[width=7cm]{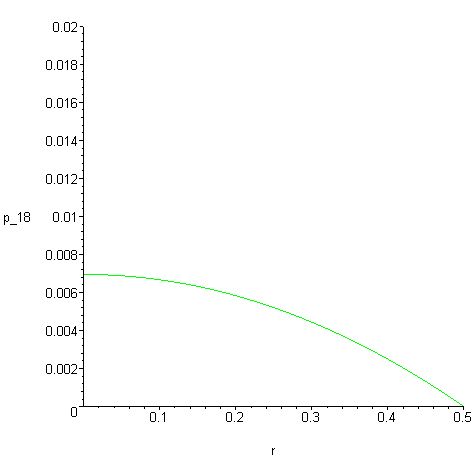}
\caption{\label{fig:pressure14} Blow up of pressure plot for the cases $n=11, 18$ on left and right respectively.}
\end{figure}


Using the expression for the density $\rho (r)$ in Eqn (\ref{eqn_25}),  let us introduce a new function, defined as:
\begin{eqnarray}
\eta (\rho )=\frac{1}{4}(n-3)\Bigr\{\sqrt{1+\frac{16\rho}{\alpha (1+K)(n-2)(n-3)^{2}}}-1 \Bigl\}
\label{eqn_32}
\end{eqnarray}
In terms of this  function we can express the $z$ variable, like $z(r)\to z(\rho )$ which simply looks as:
\begin{eqnarray}
z(r)\to z(\rho )=\frac{1}{2}\Bigr(1-\sqrt{\frac{(K+1)\eta(\rho)-1}{(K+1)\eta (\rho )}} \Bigl)
\label{eqn_33}
\end{eqnarray}
After these redefinitions of variables,  the equation of state, (EoS), can be written using the change of variables in the form $p=p(r)\to p=p(\rho )$
and the speed of sound $c^{2}_{S}=\frac{dp}{d\rho}$, where $\rho_{Surface}\leq\rho\leq \rho_{0}$

The equation of state is given by 
\begin{eqnarray}
p(\rho )=-\rho +\alpha(n-2)\Bigr((1+K)\eta^{2}(\rho)+\frac{1}{2}\eta(\rho)\sqrt{\frac{(K+1)\eta(\rho)-1}{(K+1)\eta(\rho)}}\Phi(\rho)\Bigl)
\label{eqn_36}
\end{eqnarray}
where $\Phi(\rho)$ is defined as:
\begin{eqnarray}
\Phi(\rho)=\frac{A\frac{dF_1}{dz}+B\frac{dF_2}{dz}}{A\cdot F_1+B\cdot F_2}|_{z(r)\to z(\rho )}.
\label{eqn_37}
\end{eqnarray}

The speed of sound is written as 
\begin{eqnarray}
c^{2}_{S}=-1 +\alpha(n-2)\Bigr(2(1+K)\eta(\rho)\frac{d\eta}{d\rho}+\frac{1}{2}\Bigr\{\eta(\rho)\sqrt{\frac{(K+1)\eta(\rho)-1}{(K+1)\eta(\rho)}}\Phi(\rho)\Bigl\}^{'}_{,\rho}\Bigl)
\label{eqn_38}
\end{eqnarray}
here $\{...\}^{'}_{,\rho}\equiv\frac{d\{...\}}{d\rho}$. We have thus obtained the equation of state and sound velocity involving $K, \alpha, n$ parameters.

\section{Discussion}

The main point we wish to make is that character of the key pressure isotropy equation for the BVT ansatz for the static spherically symmetric metric remains unaltered when we go higher in dimension. That means solution would always have the same form. Further it turns out that corresponding to every $4$-dimensional solution with a given $K_4$, there exists a similar solution (since $K$ is different in two cases,  solution would not be the same but similar) in higher $n$ dimension with corresponding  
$K_n = (K_4+4-n)/(n-3)$. The $4$-dimensional solution is carried over to higher dimensions but it represents different distribution. That is, the same metric  with different $K$ value describes different distributions in different dimension. For a given $K_4$, we can only go upto $n=K_4+4$ dimension when $K_n$ vanishes indicating constant density. Beyond that $K_n$ would turn negative implying density increasing with radius which is not physically tenable. Note that for $K_4=2$, the only possible solution is constant density star in $6$ dimension with $K_6=0$ while for $K_4=7$ we could have solutions for $n=5, 7, 11$ with $K_n=3, 1, 0$ respectively. As $n$ increases $K_n$ decreases to zero marking the cutoff for $n$. 

Since density gradient is always required to be negative, star is densest when density is constant implying infinite sound velocity. This is the limiting case in evolution of a static distribution in equilibrium, and it is always described by the Schwarzshild interior solution irrespective of spacetime dimension as well as  gravitational theory, Einstein or Lovelock gravity \cite{Naresh1}. Though there exists uniform density solution  in any dimension with $K=0$ given by the Schwarzschild interior solution but that would not have a correspondence with $K_4$ solution unless $n=K_4+4$. For the BVT ansatz, we have found a correspondence between $4$ dimensional solutions with their analogues in higher dimensions. For a given value of $K_4$, there is a spectrum of $K_n$ values.

It is possible to generalize the metric ansatz to bring in rotation by considering $3$-geometry being ellipsoidal. That would be very suitable for considering fluid distribution with rotation which is a very important open problem as there exists no interior metric for a rotating Kerr black hole. This may shed some new light and perhaps lead to some new insight. Another important question for further investigation is stability analysis of these higher dimensional solutions. These would be our concerns for future study.  

\section*{Acknowledgements}
AK and AM gratefully acknowledges IUCAA for the invitation and warm hospitality which facilitated this collaboration. Partial support for this work to AK was provided by Uzbekistan Foundation for Fundamental Research project F2-FA-F-116. Authors also express their thanks to R. Tikekar for the useful
discussions and his interest in this work.


\begin{thebibliography}{99}

\bibitem{Naresh1} N. Dadhich, A. Molina and A. Khugaev,{\it Phys. Rev. }{\bf D81}, 104026, (2010)
\bibitem{dhm} N Dadhich, S Hansraj, S Maharaj, Universality of isothermal fluid spheres in Lovelock gravity, arxiv:1510.07490 
\bibitem{Buchdahl}H. A. Buchdahl {\it Phys. Rev.}{\bf  116}, 1027 (1959) \& {\it Class. Quantum. Grav.}{\bf 1}, 301 (1984)
\bibitem{Vaidya} P. C. Vaidya, Ramesh Tikekar,{\it J. Astrophys. Astr.}{\bf 3}, 325 (1982).
\bibitem{Naresh} S. Mukherjee, B.C. Paul, and N. Dadhich, {\it Class. Quantum Grav.}, {\bf 14}, 3475, (1997).
\bibitem{knutsen} H. Knutsen, {\it Astrophys. Space Science} {\bf 98}, 207 (1984) 
\bibitem{Koikawa1} Koikawa T., {\it Phys. Lett. A}, {\bf 117} (1986) 279
\bibitem{Koikawa2} Koikawa T. and Yoshimura M., {\it Prog. Theor. Phys.}, {\bf 75} (1986) 977.
\bibitem{Krori1} Krori K. D., Borgohain P. and Das K., {\it Phys. Lett. A}, {\bf 132} (1988) 321.
\bibitem{Krori2} Krori K. D., Borgohain P. and Das K.,  {\it Can. J. Phys.}, {\bf 67} (1989) 25.
\bibitem{Maharaj} Maharaj S. D. and Patel L. K.  {\it Nuovo Cimento B}, {\bf 111}, (1996) 1005 
\bibitem{Patel} Patel L. K.,  Mehta N. P. and Maharaj S. D., {\it Nuovo Cimento B}, {\bf 112}, (1997) 1037.
\bibitem{Sudan} S. D. Maharaj, B. Chilambwe, S Hansraj, Phys. Rev. {\bf D91}, 084049 (2015); arxiv:1512.08072

\bibitem{Iwao} Iwao Sugai, {\it The American Mathematical Monthly}, {\bf v. 67}, No 2, 134, (1960)
\bibitem{Gauss} Morse P.M. and Feshbach H., {\it Methods of Theoretical Physics}, {\bf v.1}, New-York: McGraw-Hill, p.782,(1953).
\bibitem{Slater} Slater L.J.,{\it Generalized Hypergeometric Functions}, Cambridge University Press, p274,(1966).
\bibitem{Kanti} P. Kanti, T. Pappas and N. Pappas, arXiv: 1409.8664v1 [hep-th], 30 Sep., (2014)
\bibitem{Tangherlini} F. Tangherlini, {\it Nuovo Cimento}, {\bf 27}, 636, (1963)

\end{thebibliography}
\end{document}